\documentclass[11pt]{article}

\usepackage[]{acl}

\usepackage{times}
\usepackage{textgreek}
\usepackage{latexsym}
\usepackage{amsmath}
\usepackage{arydshln}
\usepackage{enumitem}

\usepackage{tikz}

\usepackage[T1]{fontenc}

\usepackage[utf8]{inputenc}

\usepackage{microtype}

\usepackage{graphicx}
\usepackage{tabularx}
\usepackage{booktabs}
\usepackage{multirow}
\usepackage{array}
\usepackage{makecell}
\usepackage{xcolor}
\usepackage{xstring}
\usepackage{color}
\usepackage{subcaption}
\usepackage{colortbl}

\definecolor{bittersweet}{rgb}{1.0, 0.44, 0.37}

\newcommand{\modelname}{SPAR}

\newcommand\blfootnote[1]{%
  \begingroup
  \renewcommand\thefootnote{}\footnote{#1}%
  \addtocounter{footnote}{-1}%
  \endgroup
}
\title{SPAR: Personalized Content-Based Recommendation via Long Engagement Attention}

\author{Chiyu Zhang$^{\xi, \lambda, \star}$ ~~~Yifei Sun$^{\lambda}$ ~~~Jun Chen$^{\lambda}$ ~~~Jie Lei$^{\lambda}$ ~~~\textbf{Muhammad Abdul-Mageed}$^{\xi,\gamma}$  \\ ~~~\textbf{Sinong Wang}$^{\lambda}$~~~ \textbf{Rong Jin}$^{\lambda}$ ~~~ \textbf{Sem Park}$^{\lambda}$ ~~~ \textbf{Ning Yao}$^{\lambda}$ ~~~ \textbf{Bo Long}$^{\lambda}$ \\ 
  $^{\xi}$
  The University of British Columbia~~~  $^{\lambda}$ Meta AI ~~~
  $^{\gamma}$Department of NLP \& ML, MBZUAI\\ 
  \tt chiyuzh@mail.ubc.ca \tt \{sunyifei,junchen20,jielei\}@meta.com} 

\begin{document}
\maketitle

\begin{abstract}
Leveraging users' long engagement histories is essential for personalized content recommendations. The success of pretrained language models (PLMs) in NLP has led to their use in encoding user histories and candidate items, framing content recommendations as textual semantic matching tasks.
However, existing works still struggle with processing very long user historical text and insufficient user-item interaction. In this paper, we introduce a content-based recommendation framework, SPAR, which effectively tackles the challenges of holistic user interest extraction from the long user engagement history. It achieves so by leveraging PLM, poly-attention layers and attention sparsity mechanisms to encode user's history in a session-based manner. The user and item side features are sufficiently fused for engagement prediction while maintaining standalone representations for both sides, which is efficient for practical model deployment. Moreover, we enhance user profiling by exploiting large language model (LLM) to extract global interests from user engagement history. Extensive experiments on two benchmark datasets demonstrate that our framework outperforms existing state-of-the-art (SoTA) methods.
\end{abstract}
~\blfootnote{ $^{\star}$ {Work done during Meta internship.}}

\section{Introduction}
With the prosperity of the digital world, billions of users engage daily with various digital content, including news, social media posts, and online books. Content-based recommendation systems~\cite{gu-2016-learning, embedding-2017-okura, malkiel-2020-recobert, mao_unitrec_2023} utilize textual contents (e.g., news, books) and the textual sequence of a user's engagement history to recommend more precise, relevant, and personalized content. News services, such as Google News\footnote{\url{https://news.google.com/}} 
and MSN,\footnote{\url{https://www.msn.com/en-us/news}} provide personalized news articles based on a user's browsing history.  
Platforms like Reddit\footnote{\url{https://www.reddit.com/}} and X (formerly Twitter)\footnote{\url{https://twitter.com/}} enable users to explore and engage with posts and threads of interest. 
Goodreads offers book recommendations based on a user's review history.\footnote{\url{https://www.goodreads.com/}}

Content-based recommendation systems can alleviate long-tailed and cold-start problems in traditional ID-based recommendation systems by learning the transferable semantic meanings of text content~\cite{wu_empowering_2021, liu_once_2023}. Hence, a crucial component in a content-based recommendation system is the content encoder, employed to encode textual content into latent space and capture useful information for the task. Early works use pretrained word embeddings to initialize the embedding layer and train newly initialized convolutional neural networks (CNN)~\cite{neural-2019-an}, recurrent neural networks (RNN)~\cite{embedding-2017-okura}, or attention layers~\cite{wu-2019-neural} to further aggregate the contents. Recently, PLMs have revolutionized the protocol of NLP and demonstrated their superiority across a wide range of tasks \citep{DBLP:journals/corr/abs-2302-13971,DBLP:journals/corr/abs-2303-08774,DBLP:journals/corr/abs-2304-14402,DBLP:journals/corr/abs-2306-09093,DBLP:journals/corr/abs-2307-03025,DBLP:journals/corr/abs-2312-11805}. Many studies~\cite{wu-2021-newsbert, li_miner_2022, liu_perconet_2023, liu_once_2023} have incorporated PLMs into recommendation systems to encode textual inputs. While these studies have shown success in content-based recommendation, they still face challenges in encoding long user engagement histories. For a typical recommendation system like Google News, a user's engagement history usually contains >50 news. Since each piece of news has roughly 128 tokens as its brief content, compiling all engaged news into one text field can yield sequences over 5K tokens. Due to this very long sequence length, the common practice in previous works is to encode each history content separately and fuse them later. The representation of the first token (i.e., `<SOS>' token) is usually used as the sequence-level embedding, which may fail to encapsulate fine-grained information. Separately encoding historical contents alleviates the memory issue in self-attention but lacks cross-content interactions. To address this limitation, we introduce a sparse attention mechanism for encoding session-based user behavior~\cite{liu_text_2023} and employ a poly-attention (a.k.a. codebook-based attention) module for long history integration,\footnote{In this paper, the terms `codebook-based attention' and `poly-attention' are used interchangeably. } balancing comprehensive information fusion and moderate computational demands.

Recent research~\cite{qi-2022-news, li_miner_2022, xu-2023-candidateaware} explores the early fusion of user and candidate item information to improve their interactions and click-through rate (CTR) prediction accuracy, yet this results in user/candidate representations that are dependent on each other and cannot be pre-computed. Standalone user and candidate item embeddings are important for both the lightweight retrieval stage (e.g., using similarity between pre-computed embeddings) and the ranking stage, as input for pre-computed upstream features. To overcome this challenge, we propose a novel framework that ensures standalone representations for both user and candidate items while capturing fine-grained features and enriching user-item interactions.
Inspired by the concept of \textit{post-fusion}~\cite{gong-2020-recurrent, izacard_leveraging_2021}, we employ poly-attention~\cite{humeau_poly-encoders_2020} to globally aggregate user history. To further reduce computation complexity and prevent a dramatic increase in the entropy of the attention distribution~\cite{han_lm-infinite_2023} in cases of very long sequences, we introduce three attention strategies to our poly-attention module: local sliding-window attention, global attention, and random attention. To enrich the post-interaction between a user and candidate item, we implement additional poly-attention layers to obtain multiple embeddings for both user and candidate sides. Furthermore, we enhance the user history by utilizing an LLM as a user-interest profiler to extract the user's global interests from their engagement history. 

Our main contributions are summarized as follows: 
\begin{enumerate}[label={(\arabic*)}]
\item We propose a framework for \textit{post-fusion} with \textbf{S}parse \textbf{P}oly-\textbf{A}ttention for content \textbf{R}ecommendation (SPAR),
that incorporates multiple \textit{poly-attention} layers and \textit{sparse attention} mechanisms to hierarchically fuse token level embeddings of session-based user history texts by PLM.~\modelname~effectively extracts user-interest embeddings from long history text sequences and enables sufficient interaction between user and candidate item(Section~\ref{sec:method}). 
\item We demonstrate the effectiveness of~\modelname~by testing on two widely used datasets. Our approach surpasses the SoTA methods, achieving a significant improvement of 1.48\% and 1.15\% in AUC scores for the MIND news recommendation and Goodreads book recommendation datasets, respectively (Section~\ref{subsec:main_res}).
\item Our extensive ablation studies demonstrate the impact of each component within our framework, offering insights into potential trade-offs for designing a content-based recommendation system (Section~\ref{subse:ablation}). 
\end{enumerate}

\begin{figure*}[t]
    \centering
\includegraphics[width=\linewidth]{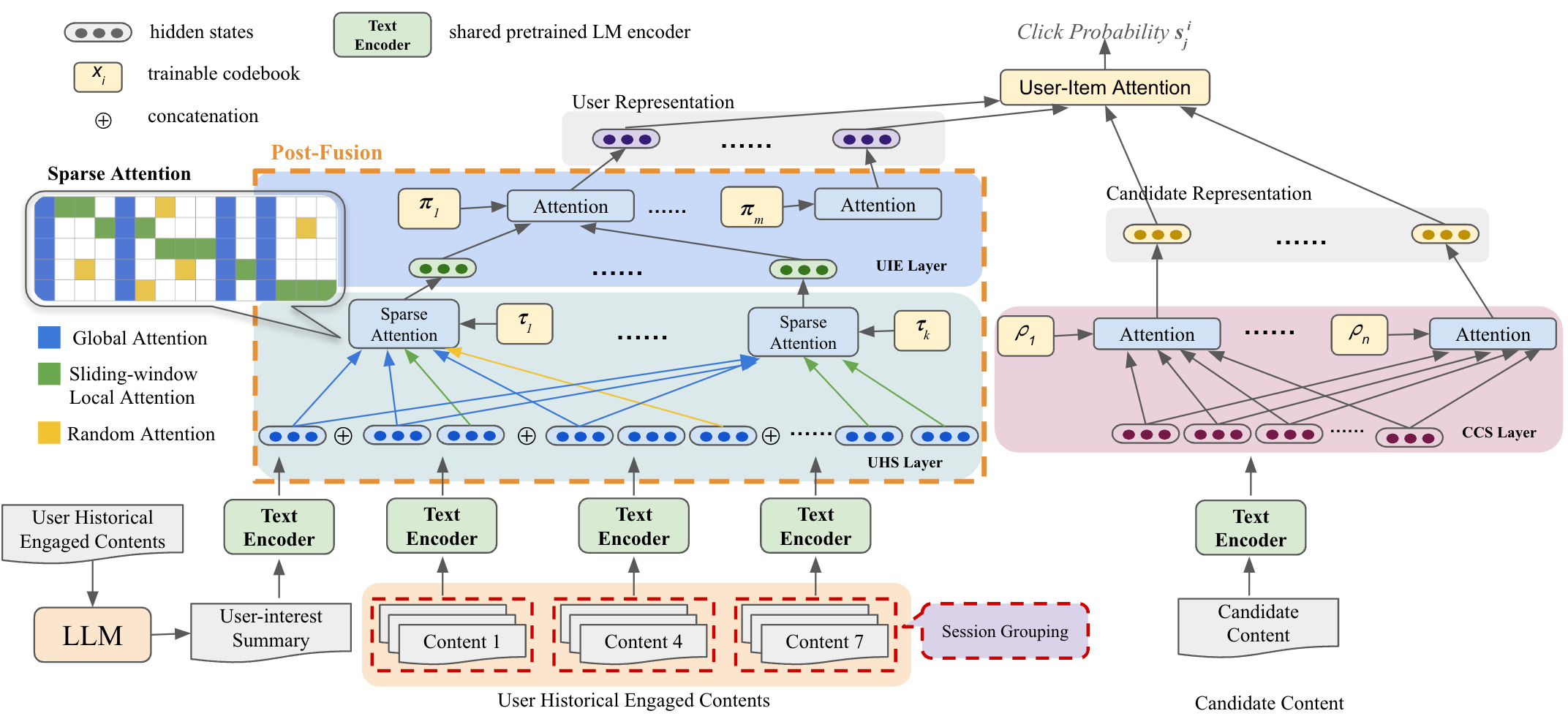}
\caption{Overview of our framework, \modelname. }
\label{fig:overview}
\end{figure*}

\section{Related Work}
Recommendation systems utilize user engagement history to identify items of potential interest to a user. Traditional recommendation systems record user engagement as a sequential list of item IDs and learn to recommend items using Markov chain assumptions~\cite{fusing-2016-he} or matrix factorization methods~\cite{factorization-2010-rendle}. However, these ID-based methods struggle with challenges like the prevalence of long-tailed items and the ``cold start" problem, where numerous new items continually emerge~\cite{methods-2002-schein, addressing-2008-lam}. Recently, several studies~\cite{cui_m6-rec_2022, li_miner_2022, liu_text_2023} have sought to mitigate these issues by introducing content-based recommendations. In these systems, user and candidate item features are described in text, this makes the recommendation systems consume pure textual features as input. The user and candidate representations are learned by various neural networks, including CNNs~\cite{neural-2019-an}, RNNs~\cite{embedding-2017-okura, wang-2022-news}, and attention mechanisms~\cite{npa-2019-wu, neural-2019-wu, qi_hierec_2021}. 

Early content-based recommendation models~\cite{neural-2019-wu,neural-2019-an} employed pretrained word embeddings, such as GloVe~\cite{pennington-2014-glove}, to initialize their embedding weights. Recent studies have been significantly bolstered by the use of PLMs. \citet{wu_empowering_2021} introduced a BERT-based two-tower framework to encode user history and candidate content, where the cosine similarity between <user, candidate> embedding pair is used as relevance score. \citet{bi_mtrec_2022} enhanced the encoder's capacity to extract effective information from news by introducing two auxiliary tasks: category classification and named entity recognition. \citet{liu_boosting_2022,liu_only_2023} did further in-domain pretraining to adapt the generic pretrained encoders. \citet{li_miner_2022} argued that a single representation vector is insufficient for capturing a user's diverse interests, and thus introduced a codebook-based attention network to obtain multiple user interest vectors. However, these studies typically encode content items individually and rely solely on the hidden state of the `[CLS]' token to represent each item. We hypothesize that this approach lacks fine-grained token-level signals which are crucial to digest the complex interaction among long sequence of content engagement history, as we show in our ablation studies (Section~\ref{subse:ablation}).

On another aspect, several studies~\cite{qi-2022-news, li_miner_2022, xu-2023-candidateaware} have introduced candidate-aware user encoder to enhance interaction and fusion between the user-side and candidate-side. \citet{li_miner_2022} introduced a category-aware attention weighting strategy that re-weights historical news based on their category similarity to the candidate news. \citet{mao_unitrec_2023} proposed an encoder-decoder framework, encoding the historical contents with a BART encoder~\cite{lewis-2020-bart}, using the candidate content as input for the decoder, and measuring the user-item relevance score by calculating the perplexity of the candidate content at the decoder. However, these approaches do not support pre-computation of user embeddings due to their dependence on awareness of the candidate item. Real-world applications require maintaining standalone representations for both users and candidate items to handle billion-scale daily interactions and support large-scale retrieval. Hence, our~\modelname~is designed to maintain independent representations, while extracting fine-grained feature sets from long history sequence with computational efficiency.

\section{Methodology}\label{sec:method}
We now introduce our proposed framework,~\modelname. As illustrated in Figure~\ref{fig:overview},~\modelname~incorporates session-based attention sparsity to encode a user's long history using a PLM. We then employ a user history summarizing (UHS) layer followed by a user interest extracting (UIE) layer to derive a comprehensive user representation (Section~\ref{subsec:user_encoder}). To further enrich the user-side representation, we utilize an LLM to generate summaries of user interests based on their engagement history (Section~\ref{subsec:user_encoder}). For candidate content, we apply the shared PLM and candidate content summarizing (CCS) layer to obtain its embeddings (Section~\ref{subsec:item_encoder}). In the end, a lightweight attention layer is used to interact user and candidate content embeddings for CTR prediction (Section~\ref{subsec:predictor}).

\subsection{Problem Formulation}
We focus on content-based recommendation tasks that solely utilize natural language text as input. Given a user $u_i$ and a candidate content (e.g., news or book) $\bar{e}_j$, our objective is to derive a relevance score $s^i_j$, indicating how likely the user $u_i$ will engage with (e.g., click) the content $\bar{e}_j$. Considering a set of candidate contents $C = \{\bar{e}_1, \bar{e}_2, \dots, \bar{e}_j\}$, these contents are ranked based on their relevance scores $\{s^i_1, s^i_2, \dots, s^i_j\}$ for user $u_i$, with the high scored contents displayed to user $u_i$ in the service. Therefore, it is critical to effectively extract user interests from their engagement history. The user $u_i$ is characterized by a sequence of $k$ historically engaged contents (such as browsed news or positively rated books) $E_{u_i} = \{e^i_1, e^i_2, \dots, e^i_k\}$, sorted descendingly by engagement time. Each content $e$ comprises text information, including title, abstract, and category. In practical applications, it is expected to pre-compute standalone embeddings for both user $u_i$ and candidate content $\bar{e}_j$ prior to calculating the relevance score. The embeddings of user $u_i$ and candidate content $\bar{e}_j$ are represented as $\Gamma_i \in \mathrm{R}^{m\times d}$ and $\Lambda_i \in \mathrm{R}^{n \times d}$, respectively, where $m, n \ll d$ and $d \ll J$, with $J$ representing the total number of candidate contents.

\subsection{User History Encoder}\label{subsec:user_encoder}
\paragraph{\textbf{Session-Based PLM Encoding}} For each content, we combine its multiple fields into a single text sequence. For instance, a news item is compiled using the template: ``News Title: $\left<title\right>$; News Abstract: $\left<abstract\right>$; News Category: $\left<category\right>$". On the user side, we concatenate the engagement history \(E = \{e_1, e_2, \dots, e_k\}\) into a long sequence, adding a start-of-sentence `$\left<SOS\right>$' and an end-of-sentence `$\left<EOS\right>$' token at the beginning and end of each content, respectively, to denote the content item boundary.
To obtain representation embeddings for users and candidate items, we employ a PLM $\Phi(\cdot)$ for encoding. The encoder is shared between the user and the candidate sides. However, the user-side content history is often much longer (exceeding 5K tokens) than the maximum length capacity of PLMs (e.g., 512 tokens for RoBERTa~\cite{liu-2019-roberta}), leading to a substantial memory requirement due to the quadratic complexity of self-attention. To mitigate this, we employ an \textit{attention sparsity mechanism}~\cite{liu_text_2023}, encoding session-based user behavior by grouping the user engagement history into $g$ subsequences $E = \{\eta_1, \eta_2,\dots, \eta_g\}$. Each session $\eta_1$ contains all $l$ tokens from $p$ contents, i.e., $\eta_1=\{e_1, e_2, \dots, e_p\} = \{w_1, w_2,\dots, w_l\}$, which can reflect the user's interests in a specific period and enhance in-session interactions. These subsequences are encoded separately by encoder $\Phi$:
\begin{equation}
    h^g_\cdot = \Phi(\eta_g),
\end{equation}
where $h^g_{\cdot} \in \mathrm{R}^{l\times d}$, $d$ is the hidden dimension of encoder $\Phi$, and $h^g_{1}$ corresponds to the hidden states of input token $w_1$ in $\eta_g$. The attention sparsity avoids the self-attention calculations between different sessions in Transformer encoder, thus lowering the computational complexity of encoding from $O(n^2)$ to $O(n)$.
 In previous work~\cite{li_miner_2022,wu_empowering_2021}, the hidden state of the first token was utilized as the representation for the entire sequence. However, this approach may not be sufficient to capture fine-grained information pertinent to user interests for a long sequence. Therefore, we take token level embedding in each subsequence and then concatenate the hidden states of all subsequences, $H = \left[h^1_{\cdot} \oplus h^2_{\cdot} \oplus \dots \oplus h^g_{\cdot} \right]$ where $H\in \mathrm{R}^{L\times d}$ and $L$ is the total number of tokens in engaged contents of a user, which can be more than 5K tokens. 

\paragraph{\textbf{LLM User Interest Summary}} Recently, LLMs~\cite{touvron-2023-llama2, DBLP:journals/corr/abs-2312-11805, wang-2024-small, DBLP:journals/corr/abs-2304-14402}  have demonstrated remarkable capabilities in general question answering, long-sequence summarization, and reasoning. In light of this, we enrich user history by employing LLM as user-interest summarizers to extract a user's global interests. We prompt LLM to generate summaries of user interests in natural language, drawing on their engagement history. These generated summaries, denoted as $\eta_+$, are then encoded by encoder $\Phi$. We prepend the sequence of hidden states $h_{\cdot}^+$ to $H$, resulting in an augmented sequence $H^+ \in \mathrm{R}^{L^+\times d}$, where $L^+$ represents the total length of the sequence after adding LLM's user-interest summary.

\paragraph{\textbf{User History Summarizing (UHS)}} While session-based encoding enables the model to capture a user's local interests within a specific period, characterizing global interests across sessions is also crucial. Inspired by the concept of post-fusion~\cite{izacard_leveraging_2021}, we propose employing a poly-attention layer~\cite{humeau_poly-encoders_2020} to globally summarize $k$ user-engaged contents into $k$ embeddings using $H^+$. This enables a comprehensive representation of user history, integrating both local and global perspectives. We learn $k$ context codes (i.e., $\tau_1, \tau_2, \dots, \tau_k$), where each $\tau_a\in \mathrm{R}^{1\times p}$ is designed to learn a contextual representation $y_a$ by attending over all $L^+$ tokens in $H^+$. The hidden dimension of code $\tau_a$ is $p$, which is smaller than $d$ to reduce compute cost. Each user-engaged content embedding $y_a$ is computed as follows:  
\begin{equation}
    y_a = W^{\tau_a} H^+, \label{eq:poly-attention_sum}
\end{equation}
where $W^{\tau_a} \in \mathrm{R}^{1 \times L^+}$ are attention weights across $L^+$ tokens in $H^+$ and are calculated as follows: 
\begin{equation}
    W^{\tau_a} = \mathrm{softmax}\left[\tau_a \mathrm{tanh}(H^+ W^c)^\top\right], \label{eq:poly-attention}
\end{equation}
where $\tau_a$ and $W^c\in \mathrm{R}^{d\times p}$ are trainable parameters. 

\paragraph{\textbf{Sparse Attention in UHS}} While the introduction of codebook queries significantly reduces the size of queries, the exceedingly long sequence of keys can still lead to the entropy of attention distribution increasing dramatically \cite{han_lm-infinite_2023} and significant compute complexity. To address this, we constrain each context code $\tau_a$  to attend only to a limited range of tokens.
Specifically, each context code $\tau_a$ attends to a subset of $\hat{\mathcal{L}}$ tokens, rather than all $L^+$ tokens, where $\hat{\mathcal{L}} \ll L^+$. Inspired by \citet{bigbird-2020-zaheer}, we implement this with three attention mechanisms, namely, \textit{local window attention}, \textit{global attention}, and \textit{random attention}, to capture both local and global information from the user's engagement history. For local attention, we set a sliding window that specifies an attention range for each context code. To provide global information across the whole user history, we make the positions of all `$\left<SOS\right>$' tokens globally visible. Additionally, we randomly select 10\% of the remaining tokens to be visible to code $\tau_a$.

\paragraph{\textbf{User Interests Extracting (UIE)}} Upon obtaining these $k$ representations of user interaction history, we concatenate them as $Y = [y_1 \oplus y_2 \oplus \dots \oplus y_k]$, and then apply another poly-attention layer to extract users' interests. Specifically, we introduce $m$ context codes (i.e., $\pi_1, \pi_2, \dots, \pi_m$) to represent the overall interests of a user as $m$ $d$-dimensional vectors. Each user-interest vector $\psi_a$ is calculated as follows:
\begin{equation}
 \begin{aligned}
    \psi_a &= \mathrm{softmax}\left[\pi_a\mathrm{tanh}(YW^f)^\top\right] Y,
\end{aligned}\label{eq:polyattention}
\end{equation}
where $\pi_a \in \mathrm{R}^{1\times p}$ and $W^f\in \mathrm{R}^{d\times p}$ are trainable parameters. We then combine $m$ user interest vectors to $\Gamma\in \mathrm{R}^{m\times d}$ as the user-side representation. 

\subsection{Candidate Content Encoder}\label{subsec:item_encoder}
To encode a candidate content, we utilize the shared encoder $\Phi$. The candidate item is constructed using the same template as outlined in Section~\ref{subsec:user_encoder}. 

\paragraph{\textbf{Candidate Content Summarizing (CCS)}} Differing from previous studies~\cite{li_miner_2022, wang-2022-news} that represent candidate content solely by the first token of the sequence, we introduce $n$ context codes (i.e., $\rho_1, \rho_2, \dots, \rho_n$) to generate multiple representations for a candidate content. Intuitively, we believe that multiple embeddings can enhance the user-candidate interactions when calculating the relevance score $s^i_j$ between user $u_i$ and candidate content $\bar{e}_j$. Similarly, we compute each candidate content vector $\omega_a$ using Eq.~\ref{eq:polyattention} with a trainable parameter $W^o$. The obtained $n$ candidate content vectors are combined into $\Lambda \in \mathrm{R}^{n \times d}$.

\subsection{Engagement Predictor}\label{subsec:predictor}
To compute the relevance score $s^i_j$, we first calculate the matching scores between the user representation embedding $\Gamma_i$ and the candidate content representation embedding $\Lambda_j$ by the inner product:
\begin{equation}
    K^i_j = \mathrm{FLATTEN} (\Gamma_i^{\top} \Lambda_j). 
\end{equation}
The matrix is flattened into a $m \times n$ dimensional vector, denoted as $K^i_j \in \mathrm{R}^{mn}$. An attention layer is applied to aggregate these $m \times n$ matching scores:
\begin{equation}
\begin{aligned}
    s^i_j &= W^p K^i_j, \\
    W^p &= \mathrm{softmax}\left[\mathrm{FLATTEN}\left[\Gamma\mathrm{gelu}(\Lambda W^s)^{\top}\right]\right],
\end{aligned}
\end{equation}
where $W^s\in \mathrm{R}^{d\times d}$ is a trainable parameter, $W^p\in\mathrm{R}^{mn}$ is the attention weights after flatten and $\mathrm{softmax}$, and $s^i_j$ is a scale of relevance score. We adhere to the common practice of training the model end-to-end using the noise contrastive estimation (NCE) loss~\cite{wu_empowering_2021,liu_once_2023}.

\section{Experiments}

\noindent\textbf{Dataset.} 
We utilize two public datasets for content-based recommendation. The first is MIND dataset~\cite{wu_mind_2020}, which comprises user behavior logs from Microsoft News. We employ the small version of MIND dataset. The second dataset is a book recommendation dataset sourced from Goodreads~\cite{wan-2018-item}, where user behaviors are inferred from book ratings.  
We provide more details and statistics of these datasets in Section~\ref{append:dataset} in Appendix.

\noindent\textbf{Baselines.} We compare our~\modelname~with several widely used and SoTA neural network-based content recommendation methods. These include methods that train text encoders from scratch, such as (1) NAML~\cite{wu-2019-neural}, (2) NRMS~\cite{neural-2019-wu}, (3) Fastformer~\cite{wu_fastformer_2021}, (4) CAUM~\cite{qi-2022-news}, and (5) MINS~\cite{wang-2022-news}, as well as systems that leverage PLMs, including (6) NAML-PLM, (7) UNBERT~\cite{zhang_unbert_2021}, (8) MINER~\cite{li_miner_2022}, and (9) UniTRec~\cite{mao_unitrec_2023}. For additional details on these baselines and their implementations, refer to Section~\ref{append:baselines} in Appendix.

\noindent\textbf{Metrics.} We adopt diverse metrics to evaluate content-based recommendation systems. These include the classification-based metric AUC~\cite{fawcett-2006-pattern}, ranking-based metrics MRR~\cite{voorhees-1999-trec8} and nDCG@top$N$ (with top$N$ = 5, 10)~\cite{jarvelink-2002-cumulated}. The Python library TorchMetrics~\cite{detlefsen-2022-torchmetrics} is employed for metric calculations. We use AUC to determine the best model on Dev set and report Test performance on all the metrics.

\noindent\textbf{Generating User-Interest Summary via LLM.}
As introduced in Section~\ref{subsec:user_encoder}, we utilize LLM as a user-interest profiler to capture global user interests. In our experiments, we employ an open-source conversational model, LLaMA2-Chat-70B~\cite{touvron-2023-llama2}, to generate concise summaries reflecting users' engagement histories. The methodology for generating user-interest summaries is elaborated in Section~\ref{append:llm_generate} in Appendix. 



\noindent\textbf{Implementation and Hyperparameters.} 
We utilize the pretrained RoBERTa-base model~\cite{liu-2019-roberta} as our content encoder. We perform hyperparameter tuning on the learning rate, the sizes of user-side and candidate-side codebooks (i.e., the UHS and CCS layers), and the local window size in the sparse attention mechanism of the UHS layers. The optimal codebook size of CCS layers is 4 and the optimal local attention window size is 512 for MIND dataset. For the Goodreads dataset, these are 4 and 256, respectively. The negative sampling ratios for the MIND and Goodreads datasets are 4 and 2, respectively. 
For both datasets, we incorporated the latest 60 user engagement contents as the user's history. The dimension size for both user and item representations is set at 200 across all experiments. Further details about hyperparameters are in Section~\ref{append:hyperparameter} of Appendix.

\begin{table}[t]
\centering
\small
\setlength\tabcolsep{4pt}
\begin{tabular}{@{}lcccc@{}}
\toprule
\multicolumn{5}{c}{\textbf{MIND-small}}                                         \\ \midrule
            & AUC            & MRR            & nDCG@5         & nDCG@10        \\  \midrule
NAML        & 66.10          & 34.65          & 32.80          & 39.14          \\
NRMS        & 63.28          & 33.10          & 31.50          & 37.68          \\
Fastformer  & 66.32          & 34.75          & 33.03          & 39.30          \\
CAUM        & 62.56          & 34.40          & 32.88          & 38.90          \\
MINS        & 61.43          & 35.99          & 34.13          & 40.54          \\\cdashline{1-5}
NAML-PLM    & 67.01          & 35.67          & 34.10          & 40.32          \\
UNBERT      & \underline{71.73}          & 38.06          & \underline{36.67}          & \underline{42.92}          \\
MINER       & 70.20          & \underline{38.10}          & 36.35          & 42.63          \\
UniTRec     & 69.38          & 37.62          & 36.01          & 42.20          \\\cdashline{1-5}
SPAR (ours) & \textbf{73.21} & \textbf{39.51} & \textbf{37.80} & \textbf{44.01} \\ \midrule
\multicolumn{5}{c}{\textbf{Goodreads}}                                          \\ \midrule
NAML        & 59.35          & 72.16          & 53.49          & 67.81          \\
NRMS        & 60.51          & 72.15          & 53.69          & 68.03          \\
Fastformer  & 59.39          & 71.11          & 52.38          & 67.05          \\
CAUM        & 55.13          & 73.06          & \underline{54.97}          & \underline{69.02}          \\
MINS        & 53.02          & 71.81          & 53.72          & 68.00          \\\cdashline{1-5}
NAML-PLM    & 59.57          & 72.54          & 53.98          & 68.41          \\
UNBERT      & \underline{61.40}          & \underline{73.34}          & 54.67          & 68.71          \\
MINER       & 60.72          & 72.72          & 54.17          & 68.42          \\
UniTRec     & 60.00          & 72.60          & 53.73          & 67.96          \\\cdashline{1-5}
SPAR (ours) & \textbf{62.55} & \textbf{73.97} & \textbf{55.48} & \textbf{69.51} \\ \bottomrule
\end{tabular}
\caption{Comparison of Test performance. The best-performing results are highlighted in \textbf{bold} font. The second-best performing model is \underline{underscored}.}\label{tab:results}
\end{table}

\section{Results}\label{sec:result}
\subsection{Overall Results}\label{subsec:main_res}
Table~\ref{tab:results} presents the Test results for models with median performance across three runs using different seeds. NAML-PLM~\cite{neural-2019-wu}, UNBERT~\cite{zhang_unbert_2021}, MINER~\cite{li_miner_2022} and UniTRec~\cite{mao_unitrec_2023}, which employ PLMs for content encoding, demonstrate a significant advantage over models that solely depend on pretrained word embeddings. For instance, we observe that NAML achieves 0.91 AUC and 0.22 AUC improvement on MIND and Goodreads datasets, respectively, when a PLM is used as the content encoder instead of training an encoder from scratch. This superiority is observed consistently across both the MIND and Goodreads datasets, underscoring the efficacy of using advanced LMs in content-based recommendation systems.
Our proposed~\modelname~achieves the highest AUC scores on both datasets, recording a remarkable 73.21 for MIND and 62.55 for Goodreads. When compared to UNBERT, our~\modelname~exhibits a significant increase of 1.48 in AUC for MIND ($t$-test on AUC, $p< 0.02$) and 1.15 for Goodreads ($t$-test on AUC, $p< 0.05$). Beyond AUC, our framework attains the best performance across all other ranking-based metrics for both datasets. We also provide the mean and standard deviation across three runs in Tables~\ref{tab:average_results} and~\ref{tab:spar_main_std} in Appendix, respectively.

\subsection{Ablation Studies and Analyses}\label{subse:ablation}
\paragraph{\textbf{Ablation Studies.}} To better understand the effectiveness of our framework, we conduct ablation studies on MIND dataset, the results of which are presented in Table~\ref{tab:ablation}. 
We first remove the UHS layer from~\modelname~(row d in Table~\ref{tab:ablation}) and use the encoder output representation of `$\left<SOS\right>$' for each piece of user historical content.\footnote{Note that the sparse attention is also removed with this UHS layer.} This alteration leads to a significant performance decrease of 1.06 in AUC, underscoring the importance of the UHS layer in summarizing user history. We then retain this UHS layer but replace our proposed sparse attention mechanism with full attention applied to the entire sequences of user engaged contents (row e). This change results in a performance drop of 0.51 in AUC, further highlighting the utility of sparse attention in utilizing the UHS layer. Following~\cite{han_lm-infinite_2023}, we empirically investigate the attention weights within the UHS layer. By processing samples from Dev set through each model, we extract the attention weights for the first three codes in the UHS layer. We then calculate the entropy of the attention distribution for each code per sample, averaging across the three poly-attention codes and all samples. Our findings indicate a decrease in entropy from 7.74 to 5.66 upon implementing sparse attention in the UHS layer.

We explore the impact of removing the UIE layer by setting the query codebook size to 1 (row f), which also results in a performance decrease of 0.16 in AUC, indicating that a single user embedding is insufficient.
We then experiment with newly initializing the Transformer encoder layers and training them end-to-end (row a), which leads to a significant performance drop of 5.76 in AUC, demonstrating the crucial role of using a PLM. We also keep the PLM frozen and only train the new layers (row b), which results in a performance drop of 2.86 AUC.  
After removing session-based history content grouping (row g), the LLM-generated user-interest summary (row h), and the CCS layer (row i), we observe slight performance drops of 0.07, 0.06, and 0.02 in AUC, respectively. We then also remove these three components together (row j), but this leads to a large performance drop of 0.46 AUC. These findings suggest that while these components individually contribute to smaller performance enhancements, collectively, they are integral to the overall effectiveness of our framework.

We also experiment with using only the LLM-generated user-interest summary as the user-side input and setting the codebook size of the UHS layer to 1, considering that the user side consists of a single history content (i.e., user-interest summary). This setting (row b) results in an AUC of 70.37 and an MRR of 37.46, significantly underperforming SPAR. This indicates the importance of using the original user history for learning user interests.

\begin{table}[t]
\centering
\small
\setlength\tabcolsep{3pt}
\begin{tabular}{@{}lcccc@{}}
\toprule
                          & AUC   & MRR   & nD@5  & nD@10 \\ \midrule
\modelname                      & \textbf{73.21} & \textbf{39.51} & 37.80 & 44.01 \\ \hline
(a) wo PLM                    & 66.45 & 32.33 & 31.22 & 37.47 \\
(b) wo original history       & 70.37  & 37.46  & 35.91	 & 42.19 \\
(c) freeze PLM                & 70.85 & 38.75 & 37.21 & 43.17 \\
(d) wo UHS       & 72.15 & 38.71 & 37.38 & 43.56 \\
(e) wo attention sparsity     & 72.70 & 39.10 & 37.36 & 43.65 \\
(f) wo UIE      & 73.05 & 39.43 & 37.84 & 44.05 \\
(g) wo session grouping & 73.14 & 39.50 & 37.79 & 43.92 \\
(h) wo LLM sum.  & 73.15 & 39.40 & 37.61 & 44.01 \\
(i) wo CCS     & 73.19 & 39.46 & \textbf{38.06} & \textbf{44.07} \\ \cdashline{1-5}
(j) wo (g), (h), (i)      & 72.75 &	39.39 & 	37.55	&   43.87 \\ 
\bottomrule
\end{tabular}
\caption{Test performance of ablation studies on MIND dataset. \textbf{nD@$N$} represents nDCG@$N$.}\label{tab:ablation}
\end{table}


\begin{table}[ht]
\small
\centering
\begin{tabular}{@{}lcccc@{}}
\toprule
\multicolumn{1}{c}{\textbf{}} & AUC            & MRR            & nDCG@5         & nDCG@10        \\ \midrule
\modelname                          & \textbf{73.21} & \textbf{39.51} & \textbf{37.80} & \textbf{44.01} \\
wo UHS                        & 72.15          & 38.71          & 37.38          & 43.56          \\\hline
S-Longf.                    & 73.02          & 39.34          & 38.11          & 43.92          \\
S-BigB.                       & 71.17          & 38.13          & 36.82          & 43.05          \\ \bottomrule
\end{tabular}
\caption{Comparing different backbone models.}\label{tab:abl_long_model}
\end{table}
\begin{figure}[t]
    \centering
\includegraphics[width=\linewidth]{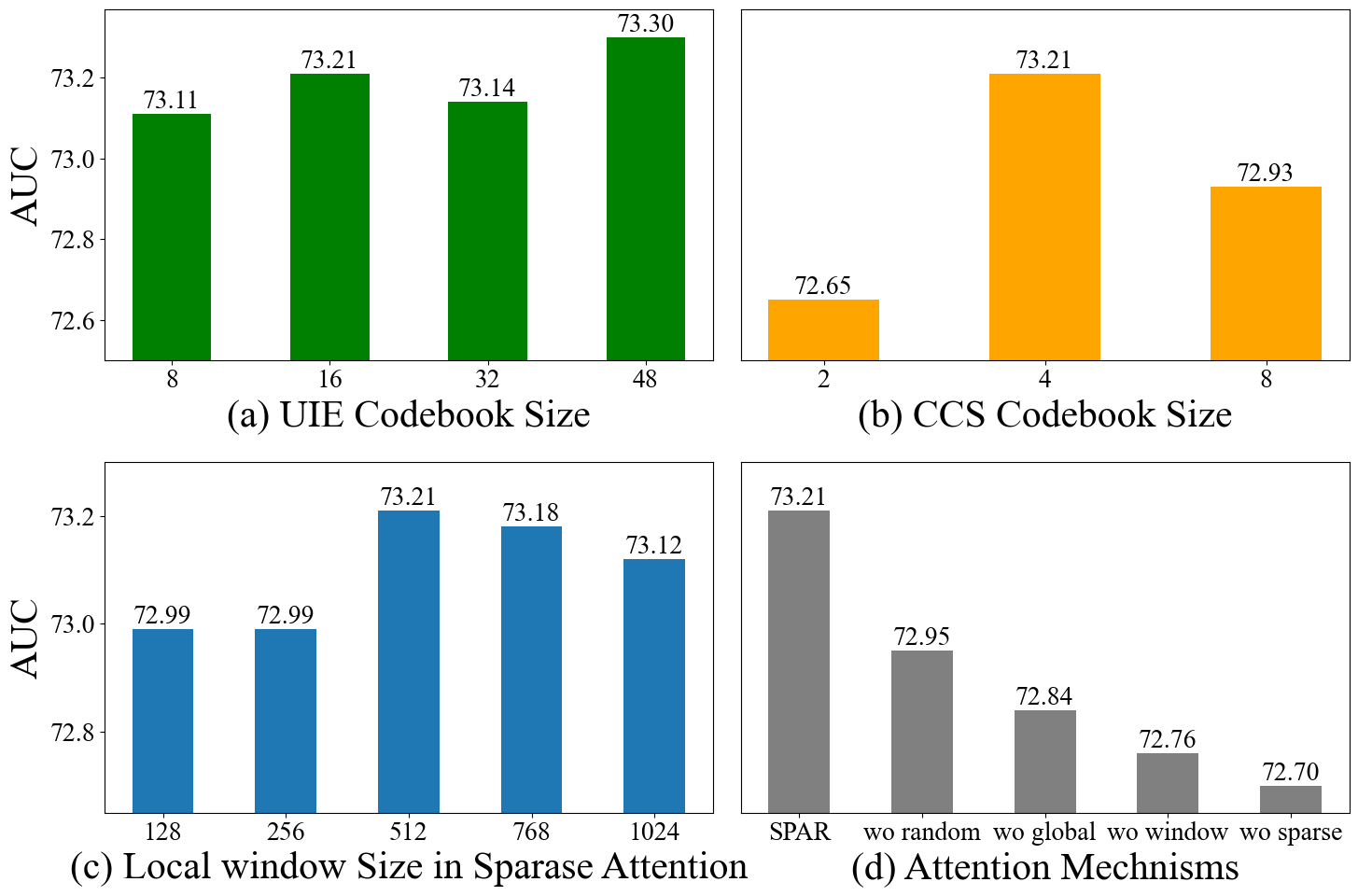}
\caption{Influence of different hyperparameters.}
\label{fig:hyperparmeters}
\end{figure}

\begin{figure}[ht]
\centering
\begin{subfigure}[]{.24\textwidth}
  \centering
  \includegraphics[width=\linewidth]{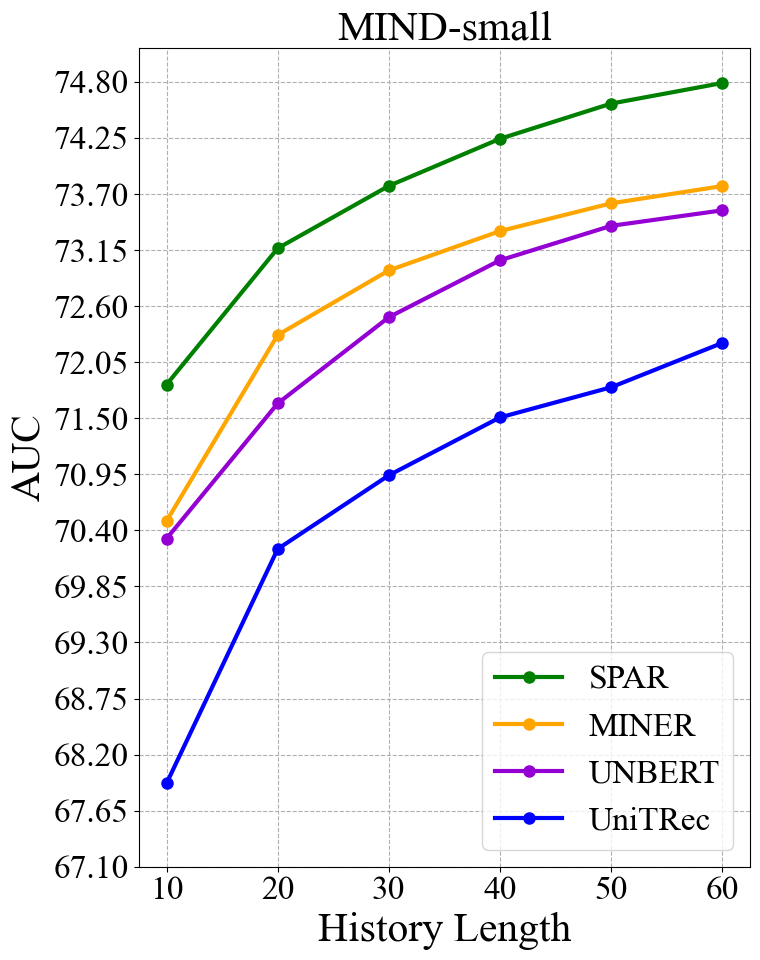}
\end{subfigure}%
\begin{subfigure}[]{.24\textwidth}
  \centering
  \includegraphics[width=\linewidth]{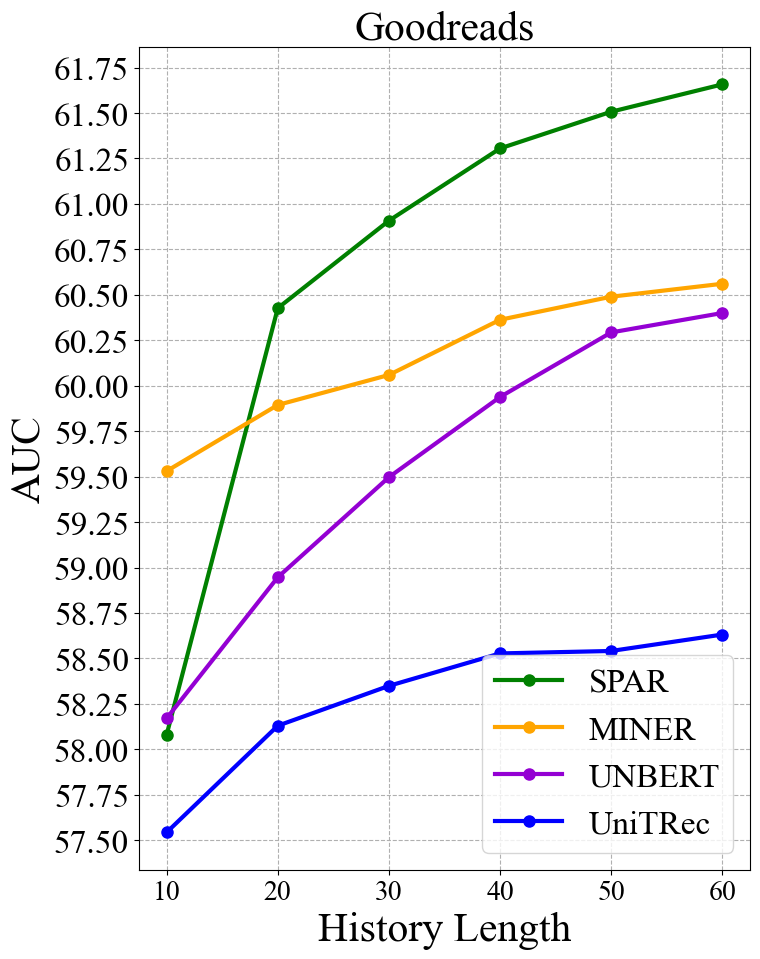}
\end{subfigure} 
\caption{Effects of user engagement history length.}\label{fig:history_length}
\end{figure}

\noindent\textbf{Handling Long Sequences.} As the aforementioned results indicate, managing long sequences of user engagement history is a crucial component. We explore whether using PLM for processing long sequences could be a superior solution compared to our proposed framework. Two long sequence PLMs are utilized in this experiment, namely, Longformer-base~\cite{beltagy_longformer_2020} and BigBird-base~\cite{bigbird-2020-zaheer}. These models are Transformer encoder-only models, pretrained for processing up to 4,096 tokens. We leverage these long sequence PLMs as a content encoder and remove the UHS layer from our framework. To ensure the encoding of all 60 historical contents of a given user, we divide user history into two groups, each potentially containing up to 30 user historical engagement contents.\footnote{A single sequence of 60 user history contents exceeds the maximal sequence length in some cases.} We extract the encoder output representation of `$\left<SOS\right>$' for each piece of user historical content and pass the sequence of user historical contents through the UIE layer in our framework. We refer to these two variants as~\modelname-Longformer and~\modelname-BigBird. As Table~\ref{tab:abl_long_model} illustrates, both~\modelname-Longformer and~\modelname-BigBird underperform our \modelname~ with a performance gap of 0.19 and 2.04 AUC, respectively. While \modelname-Longformer performs better than our ablation model with the UHS layer removed, it still lags behind \modelname.\footnote{When we use eight Nvidia A100 40G GPUs, training \modelname-Longformer is also notably slower, requiring 8.5 hours per epoch, whereas \modelname~completes in 3 hours per epoch. }
This suggests the effectiveness of our proposed method in handling long user history with a post-fusion strategy. 


\noindent\textbf{Influence of Hyperparameters.} Figure~\ref{fig:hyperparmeters} illustrates the effects of various hyperparameters. Figure~\ref{fig:hyperparmeters}a shows that generally an increase in UIE codebook size correlates with higher AUC scores. A UIE codebook size of 16 is preferred to minimize storage requirements, albeit at the cost of a marginal performance decline. Specifically, a UIE codebook size of 16 results in a 0.09 decrease in AUC but achieves a threefold reduction in storage space compared to a size of 48. Regarding the codebook size of the CCS layer (Figure~\ref{fig:hyperparmeters}b), our findings suggest that an increased number of codebooks on the candidate item side does not enhance model performance. We hypothesize that excessively numerous codebooks for a single item might introduce superfluous parameters, thereby detrimentally impacting performance. Figure~\ref{fig:hyperparmeters}c indicates that the optimal sliding window size for sparse attention in the UHS is 512, with longer attention spans (exceeding 512) in UHS codebook-based attention adversely affecting model efficacy. In Figure~\ref{fig:hyperparmeters}d, our analysis of various sparse attention mechanisms in the UHS layer reveals performance decrements of 0.26, 0.37, and 0.45 in AUC upon the removal of random, global, and local window attention mechanisms, respectively. These findings underscore the significance of incorporating both local window and global attention mechanisms in the UHS layer.

\noindent\textbf{Effects of User Engagement History Length.}
We examine the impact of the quantity of user-engaged history on model performance. We utilize a subset of Test users who are not present in the Train set and possess a minimum of 60 engaged contents. In cases where a user has multiple impressions, we randomly select one impression to represent that user. The user histories are truncated to the $K$ most recent contents, with $K = \{10, 20, 30, 40, 50, 60\}$, while maintaining the same candidate contents in the impression. As shown in Figure~\ref{fig:history_length},~\modelname~ outperforms the three strong baselines across all engagement history lengths in MIND dataset. Notably, while~\modelname~initially trails behind MINER and UNBERT with only 10 historical entries, it demonstrates marked improvement as the number of user-engaged histories increases. These findings underscore the effectiveness of extensive user histories and highlight the capability of~\modelname~in processing lengthy sequences for content-based recommendation. We provide results in other ranking metrics in Tables~\ref{tab:history_leng_mrr},~\ref{tab:history_leng_ndcg5}, and~\ref{tab:history_leng_ndcg10} in Appendix.

\begin{table}[t]
\centering
\small
\begin{tabular}{@{}llcccc@{}}
\toprule
                          & \textbf{} & \multicolumn{2}{c}{\textbf{MIND}} & \multicolumn{2}{c}{\textbf{Goodreads}} \\ \cmidrule(l){3-4}\cmidrule(l){5-6} 
                          & \textbf{} & \textbf{AUC}   & \textbf{nD@5}  & \textbf{AUC}     & \textbf{nD@5}     \\ \midrule
\multirow{4}{*}{\rotatebox[origin=c]{90}{Old User}} & UNBERT    & 71.65          & 36.98            & 61.27            & 54.82               \\
                          & MINER     & 70.37          & 36.41            & 60.76            & 54.92               \\
                          & UniTRec   & 69.87          & 36.93            & 60.08            & 53.94               \\
                          & \modelname      & \textbf{73.29}          & \textbf{38.08}            & \textbf{62.67}            & \textbf{55.42}               \\ \midrule
\multirow{4}{*}{\rotatebox[origin=c]{90}{New User}} & UNBERT    & 71.75          & 36.95            & 60.94            & 54.40               \\
                          & MINER     & 70.18          & 36.28            & 60.59            & 54.20               \\
                          & UniTRec   & 69.85          & 36.66            & 59.37            & 52.78               \\ 
                          & \modelname      & \textbf{73.19}          & \textbf{37.74}            & \textbf{62.51}            & \textbf{55.51}               \\
                          \bottomrule
\end{tabular}
\caption{Comparing performance on old and new users. The new users are users that do not occur in Train set. nD@5: nDCG@5.}\label{tab:new_users}
\end{table}

\noindent\textbf{Performance on Old and New Users.}
Test users are divided into two categories: old users, who are present in the Train set, and new users, who have not been previously included in the Train set. Table~\ref{tab:new_users} presents a comparison of model performance between these two user groups. It is observed that our~\modelname~surpasses all baselines in both groups across two datasets. Generally, the performance of the old user group is slightly better than that of the new user group. For instance, the performance gap between the old and new user groups in our~\modelname~is 0.10 in AUC. This outcome suggests the effective transferability of content-based recommendation methods to new users.

\section{Conclusion}
We introduced a novel framework,~\modelname, designed to obtain independent representations for both users and candidate content. Our~\modelname~excels at extracting fine-grained features from long user engagement histories, significantly enhancing the post-interaction dynamics between users and candidate contents. By integrating a poly-attention scheme with a sparse attention mechanism,~\modelname~effectively aggregates lengthy user history sequences while maintaining relatively low computational costs. Through our experiments on two benchmark datasets, we have demonstrated that our framework achieves SoTA performance. Our ablation studies and model analyses have demonstrated the effectiveness of each component in our proposed framework and highlighted its robustness for content-based recommendation.


\section{Limitations}
In this study, we exclusively focus on content-based recommendation, with text features as the only input. We acknowledge that real-world recommendation systems require non-content related sparse/dense features or content features from other modalities. While our framework is tailored for text content, we envision it as an effective component contributing to the overall recommendation system. 

In our experiments, we opted for a base-sized encoder model to strike a balance between computational efficiency and model performance. However, for practical real-world applications, smaller-sized models are preferable due to their faster inference speed. In future work, we plan to investigate the use of other smaller-sized models as the backbone, aiming to optimize the trade-off between performance and efficiency in real-world scenarios.

\section{Ethical Considerations}

We employed an LLM to generate summaries of user interests, which serve as input for learning user-side representations. It is critical to recognize that the LLM's outputs may reflect societal biases~\cite{lucy-bamman-2021-gender} or produce inaccuracies known as hallucinations~\cite{zhang-2023-hallucination}. Consequently, the recommendation models developed using these summaries might also exhibit such biases. However, we expect that the active research aimed at enhancing the social fairness, accuracy, and reliability of LLMs~\cite{dev-2022-measures, ji-2023-towards, zhou-etal-2023-context} will also improve the performance and ethical standards of recommendation systems incorporating LLMs as a component.

In our experiments, we utilized two publicly available datasets designated for research purposes. The datasets' original authors~\cite{wu_mind_2020, wan-2018-item} have anonymized user identities to protect privacy.

\bibliography{anthology,custom}
\bibliographystyle{acl_natbib}

\appendix
\label{sec:appendix}
\newpage
\noindent\textbf{\large Appendices}
\section{Dataset} \label{append:dataset}

We utilize two public benchmark datasets for content-based recommendation. The first is MIND dataset from \citet{wu_mind_2020}, which comprises user behavior logs from Microsoft News. This dataset includes positive and negative labels that indicate whether a user clicked on a news article in an impression. We employ the small version of the MIND dataset, containing 94,000 users and 65,238 news articles for training and validation. The validation set is divided into a 10\% Dev set and a 90\% Test set. Each news article in the MIND dataset is characterized by a title, abstract, and category label. Our second dataset is a book recommendation dataset sourced from Goodreads~\cite{wan-2018-item}, where user behaviors are inferred from book ratings. Positive labels signify ratings above 3, while negative labels indicate ratings below 3. Each book entry includes a name, author, description, and genre category. Table~\ref{tab:dataset} details the data distribution and statistics.


\begin{table*}[ht]
\centering
\small
\setlength\tabcolsep{3pt}
\begin{tabular}{@{}lrrr|rrr|lrr@{}}
\toprule
     \multicolumn{1}{c}{Dataset}        & \multicolumn{3}{c}{MIND}                                                       & \multicolumn{3}{c|}{Goodreads}                                                  & \multicolumn{1}{c}{\multirow{2}{*}{Dataset}}       & \multicolumn{1}{c}{\multirow{2}{*}{MIND}} & \multicolumn{1}{c}{\multirow{2}{*}{Goodreads}} \\ \cmidrule(lr){1-1} \cmidrule(lr){2-4}\cmidrule(lr){5-7}
   \multicolumn{1}{c}{Split}   & \multicolumn{1}{c}{Train} & \multicolumn{1}{c}{Dev} & \multicolumn{1}{c}{Test} & \multicolumn{1}{c}{Train} & \multicolumn{1}{c}{Dev} & \multicolumn{1}{c|}{Test} & \multicolumn{1}{c}{}       & \multicolumn{1}{c}{}                      & \multicolumn{1}{c}{}                           \\ \midrule
\# content   & 51,283                    & 21,352                  & 41,496                    & 309,047                   & 234,232                 & 247,242                   & \# of history/user         & 22                                        & 47                                             \\
\# users     & 50,000                    & 6,679                   & 46,549                    & 21,450                    & 16,339                  & 17,967                    & \# category                & 18                                        & 11                                             \\
\# new users & \multicolumn{1}{c}{-}     & 5,862                   & 41,020                    & \multicolumn{1}{c}{-}     & 2,930                   & 3,199                     & \# tokens/title            & 29                                        & 16                                             \\
\# positive  & 236,344                   & 10,775                  & 100,608                   & 198,403                   & 75,445                  & 93,156                    & \# tokens/abstract         & 144                                       & 190                                            \\
\# negative  & 5,607,100                 & 249,607                 & 2,380,008                 & 458,435                   & 141,977                 & 154,016                   & \# tokens/user summary & 70                                        & 106                                            \\ \bottomrule
\end{tabular}
\caption{Dataset Statistics. The row `\# new users' indicates the number of users not included in the Train set. `\# tokens/user summary' represents the average length of user interest summaries generated by LLM. The number of tokens are calculated using the RoBERTa-base model's vocabulary.}
\label{tab:dataset}
\end{table*}


\section{Baselines}\label{append:baselines}
We compare our proposed model with previous SoTA neural network-based methods for content-based recommendation:
\begin{enumerate}[label={(\arabic*)}]
    \item NAML~\cite{wu-2019-neural} trains a contents representation by using a sequence of CNN and additive attention and pretrained word embeddings. To get a user's embedding, additional additive attention is applied to the embeddings of the user's engaged contents. 
    \item NRMS~\cite{neural-2019-wu} utilizes pretrained word embeddings, multi-head self-attention, and additive attention layers to learn representations for both users and candidate content.
    \item Fastformer~\cite{wu_fastformer_2021} is an efficient Transformer architecture based on additive attention. 
    \item CAUM~\cite{qi-2022-news} enhances NRMS by incorporating title entities into content embeddings and introducing candidate-aware self-attention for user embedding.
    \item MINS~\cite{wang-2022-news} advances NRMS by implementing a multi-channel GRU-based recurrent network to learn sequential information in user history.
    \item NAML-PLM uses a PLM as a content encoder instead of a training encoder from scratch. 
    \item UNBERT~\cite{zhang_unbert_2021} utilizes a PLM to encode input content and captures user-content matching signals at both item-level and word-level.
    \item MINER~\cite{li_miner_2022} employs a PLM as a text encoder and adopts a poly attention scheme to extract multiple user interest vectors for user representation. MINER is arguably the most widely referred and top-ranked model in the MIND dataset leaderboard.\footnote{\url{https://msnews.github.io/}} 
    \item UniTRec~\cite{mao_unitrec_2023} utilizes encoder-decoder architecture (i.e., BART) and encodes user history by encoder and candidate content by decoder respectively. They rank the candidate contents based on their perplexity from the decoder and discriminative scoring head.\footnote{We use the predictions from discriminative scoring head to calculate metrics.} 
\end{enumerate}

We adhere to the optimal hyperparameters of these SoTA baselines and train as well as evaluate them on our data splits. For NAML, NRMS, Fastformer, and NAML-PLM, we exploit the implementation by~\citet{legommenders-2023-liu}. For CAUM and MINS, we use the implementation provided by \citet{iana_newsreclib_2023}. UNBERT, MINER, and UniTRec are implemented via the author-released scripts, respectively.\footnote{UNBET: \url{https://github.com/reczoo/RecZoo/tree/main/pretraining/news/UNBERT}, MINER: \url{https://github.com/duynguyen-0203/miner}, UniTRec: \url{https://github.com/Veason-silverbullet/UniTRec}.}

\section{Generating User-Interest Summary via LLM}\label{append:llm_generate}
\begin{figure}
    \centering
    \includegraphics[width=\linewidth]{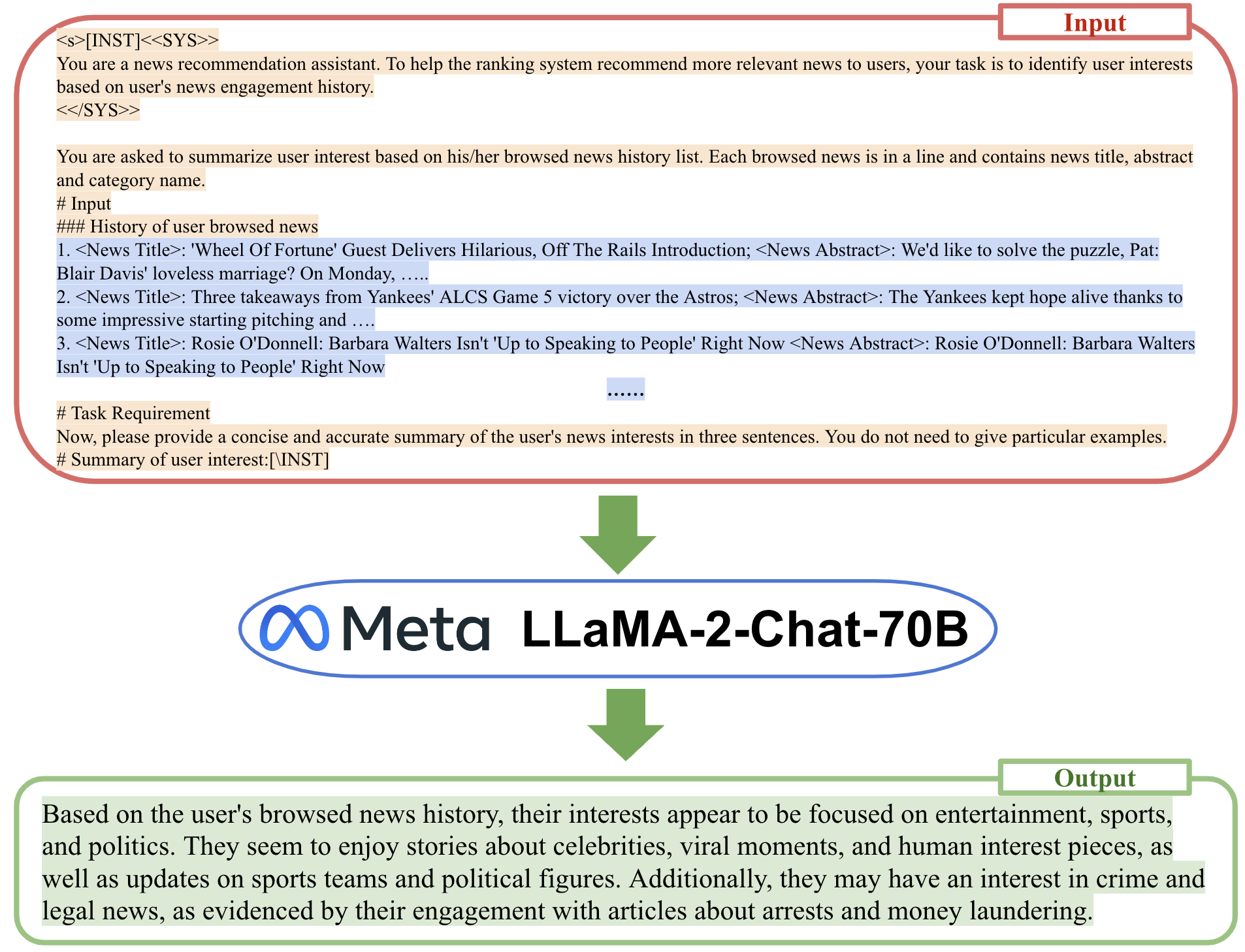}
    \caption{Example of prompting an LLM as a user profiler. The text within the red box represents the input to the LLM, while the text within the green box is the LLM's generated output. In the input text, the part highlighted in orange signifies the task instruction, and the text in blue denotes the user's browsed news history.}
    \label{fig:prompt_llm}
\end{figure}

We employ an open-source conversational LLM, specifically LLaMA2-Chat-70B~\cite{touvron-2023-llama2}, to generate summaries of user interests based on their engagement history. 
Figure~\ref{fig:prompt_llm} illustrates an example of our input and the generated output for the MIND dataset. Following the prompt template of LLaMA2-Chat models, we start with a system instruction including special tokens to establish the task context, followed by a list of user-browsed news history, sorted from the most recent to the oldest. Each listed news item includes its title, abstract, and category name. We limit the input prompt to 30 engaged contents and truncate longer news abstracts or book descriptions to 100 words to maintain sequence length within the maximal capacity of LLaMA2. The prompt concludes with a task-specific instruction requesting the model to summarize user interests in three sentences.\footnote{We manually fine-tuned this input template based on a few examples.} The lower section of Figure~\ref{fig:prompt_llm} displays a example generation from the LLM. As demonstrated in Table~\ref{tab:dataset}, the average length of these generated summaries is 70 tokens for the MIND dataset and 106 tokens for the Goodreads dataset.

\section{Implementation and Hyperparameters}\label{append:hyperparameter}
We utilize the pretrained RoBERTa-base model~\cite{liu-2019-roberta} as our backbone model for encoding contents. We set the batch size to 128 and train the model for 10 epochs using the Adam optimizer, with a linearly decaying learning rate and a 10\% warm-up phase. For the newly initialized layers, we employ a learning rate five times higher than the base learning rate. Hyperparameter tuning is performed within the following search space: base peak learning rate $ = \{5e-5, 2e-5\}$, size of user-side context codes $= \{16, 32, 48\}$, candidate-side codebook size $= \{2, 4, 8\}$, and size of the local window in sparse attention $= \{256, 512, 1024\}$. For computational efficiency, 20\% of the training data is randomly sampled for hyperparameter tuning, with the best parameters determined based on Dev set performance. The identified optimal hyperparameters for both datasets are a peak learning rate of $2e-5$ and a user-side context codebook size of 16. For the MIND dataset, the optimal candidate-side codebook size is 4 and the local attention window size is 512. For the Goodreads dataset, these are 4 and 256, respectively. For all experiments, we incorporated the most recent 60 user engagement contents as the user's history. In the MIND dataset, the negative sampling ratio is set to 4, the maximum length of a news title to 32 tokens, and the maximum length of a news abstract to 72 tokens. In the Goodreads dataset, the negative sampling ratio is 2, the maximum length of a book name is 24 tokens, and the maximum length of a book description is 85 tokens. The dimension size for both user and item representations is set at 200 across all experiments. Evaluations on the Dev set are conducted every 610 steps (one-third of an epoch) for the MIND dataset and every 520 steps for the Goodreads dataset, with the Test performance of the best model reported. All models are trained using eight Nvidia A100 40G GPUs.

\section{Results}
\begin{table}[ht]
\centering
\small
\setlength\tabcolsep{4pt}
\begin{tabular}{@{}lcccc@{}}
\toprule
\multicolumn{5}{c}{\textbf{MIND-small}}                                                                                    \\ \midrule
            & \multicolumn{1}{c}{AUC} & \multicolumn{1}{c}{MRR} & \multicolumn{1}{c}{nDCG@5} & \multicolumn{1}{c}{nDCG@10} \\ \midrule
NAML        & 65.42                   & 34.44                   & 32.57                      & 38.90                       \\
NRMS        & 63.69                   & 33.15                   & 31.33                      & 37.63                       \\
Fastformer  & 66.22                   & 34.47                   & 32.74                      & 39.12                       \\
CAUM        & 62.78                   & 34.52                   & 33.09                      & 39.22                       \\
MINS        & 61.00                   & 35.58                   & 33.86                      & 40.23                       \\\cdashline{1-5}
NAML-PLM    & 67.11                   & 35.57                   & 34.27                      & 40.41                       \\
UNBERT      & \underline{71.69}                   & 38.10                   & \underline{36.71}                      & \underline{42.92}                       \\
MINER       & 70.23                   & \underline{38.24}                   & 36.55                      & 42.82                       \\
UniTRec     & 69.48                   & 37.76                   & 36.19                      & 42.35                       \\\cdashline{1-5}
SPAR (ours) & \textbf{73.20}          & \textbf{39.34}          & \textbf{37.77}             & \textbf{44.04}              \\ \midrule
\multicolumn{5}{c}{\textbf{Goodreads}}                                                                                      \\ \midrule
NAML        & 59.18                   & 72.05                   & 53.59                      & 67.90                       \\
NRMS        & 60.55                   & 72.18                   & 53.66                      & 68.02                       \\
Fastformer  & 59.39                   & 71.18                   & 52.37                      & 67.09                       \\
CAUM        & 55.03                   & 73.31                   & \underline{55.03}                      & \underline{69.08}                       \\
MINS        & 53.03                   & 71.92                   & 53.59                      & 67.98                       \\\cdashline{1-5}
NAML-PLM    & 59.43                   & 72.52                   & 53.78                      & 68.38                       \\
UNBERT      & \underline{61.37}                   & \underline{73.38}                   & 54.64                      & 68.70                       \\
MINER       & 60.69                   & 72.87                   & 54.32                      & 68.51                       \\
UniTRec     & 59.64                   & 72.52                   & 53.51                      & 67.86                       \\\cdashline{1-5}
SPAR (ours) & \textbf{62.43}          & \textbf{73.87}          & \textbf{55.28}             & \textbf{69.42}              \\ \bottomrule
\end{tabular}
\caption{Comparison of Test performance. The average scores over three runs are reported, with the best-performing results highlighted in \textbf{bold} and the second-best \underline{underscored}.}\label{tab:average_results}
\end{table}

\begin{table}[t]
\centering
\small
\setlength\tabcolsep{4pt}
\begin{tabular}{@{}lcccc@{}}
\toprule
\multicolumn{5}{c}{\textbf{MIND-small}}      \\ \midrule
            & AUC  & MRR  & nDCG@5 & nDCG@10 \\ \midrule
NAML        & 1.65 & 0.52 & 0.58   & 0.57    \\
NRMS        & 1.12 & 1.02 & 1.04   & 0.91    \\
Fastformer  & 0.57 & 0.67 & 0.67   & 0.57    \\
CAUM        & 0.85 & 1.02 & 0.92   & 1.04    \\
MINS        & 1.13 & 0.83 & 0.67   & 0.62    \\\cdashline{1-5}
NAML-PLM    & 0.21 & 0.41 & 0.32   & 0.48    \\
UNBERT      & 0.20 & 0.37 & 0.23   & 0.21    \\
MINER       & 0.43 & 0.31 & 0.40   & 0.36    \\
UniTRec     & 0.34 & 0.40 & 0.44   & 0.35    \\\cdashline{1-5}
SPAR (ours) & 0.10 & 0.15 & 0.20   & 0.07    \\ \midrule
\multicolumn{5}{c}{\textbf{Goodreads}}       \\ \midrule
NAML        & 0.31 & 0.24 & 0.23   & 0.18    \\
NRMS        & 0.06 & 0.16 & 0.10   & 0.09    \\
Fastformer  & 0.09 & 0.25 & 0.17   & 0.14    \\
CAUM        & 0.45 & 0.55 & 0.24   & 0.29    \\
MINS        & 0.55 & 0.27 & 0.27   & 0.18    \\\cdashline{1-5}
NAML-PLM    & 0.34 & 0.04 & 0.45   & 0.09    \\
UNBERT      & 0.15 & 0.13 & 0.11   & 0.05    \\
MINER       & 0.13 & 0.48 & 0.41   & 0.35    \\
UniTRec     & 0.97 & 0.40 & 0.39   & 0.29    \\\cdashline{1-5}
SPAR (ours) & 0.29 & 0.41 & 0.36   & 0.35    \\ \bottomrule
\end{tabular}
\caption{Standard deviation of Test performance over three runs.}\label{tab:spar_main_std}
\end{table}

Table~\ref{tab:average_results} and~\ref{tab:spar_main_std} display the mean and standard deviation of the Test set performance for all baselines and SPAR.

Tables~\ref{tab:history_leng_auc}, \ref{tab:history_leng_mrr}, \ref{tab:history_leng_ndcg5}, and \ref{tab:history_leng_ndcg10} present the comprehensive results from investigating the effects of user engagement history length on AUC, MRR, ndcg@5, and ndcg@10, respectively.

\begin{table}[h]
\centering
\scriptsize
\begin{tabular}{@{}lcccccc@{}}
\toprule
\multicolumn{1}{c}{\textbf{History leng.}} & \textbf{10} & \textbf{20} & \textbf{30} & \textbf{40} & \textbf{50} & \textbf{60} \\ \midrule
\multicolumn{7}{c}{\textbf{MIND}}                                                                                              \\ \midrule
UNBERT                                     & 70.32       & 71.65       & 72.49       & 73.05       & 73.39       & 73.54       \\
MINER                                      & 70.49       & 72.32       & 72.95       & 73.34       & 73.61       & 73.78       \\
UniTRec                                    & 67.92       & 70.22       & 70.94       & 71.51       & 71.80       & 72.24       \\
SPAR                                       & \textbf{71.83}       & \textbf{73.17}       & \textbf{73.78}       & \textbf{74.24}       & \textbf{74.59}       & \textbf{74.79}       \\ \midrule
\multicolumn{7}{c}{\textbf{Goodreads}}                                                                                         \\ \midrule
UNBERT                                     & 58.17       & 58.95       & 59.50       & 59.94       & 60.29       & 60.40       \\
MINER                                      & \textbf{59.53}       & 59.89       & 60.06       & 60.36       & 60.49       & 60.56       \\
UniTRec                                    & 57.54       & 58.13       & 58.35       & 58.53       & 58.54       & 58.63       \\
SPAR                                       & 58.08       & \textbf{60.43}       & \textbf{60.91}       & \textbf{61.31}       & \textbf{61.51}       & \textbf{61.66}       \\ \bottomrule
\end{tabular}
\caption{Complete AUC results from investigating the effects of user engagement history length.}\label{tab:history_leng_auc}
\end{table}

\begin{table}[h]
\centering
\scriptsize
\begin{tabular}{@{}lllllll@{}}
\toprule
\multicolumn{1}{c}{\textbf{History leng.}} & \multicolumn{1}{c}{\textbf{10}} & \multicolumn{1}{c}{\textbf{20}} & \multicolumn{1}{c}{\textbf{30}} & \multicolumn{1}{c}{\textbf{40}} & \multicolumn{1}{c}{\textbf{50}} & \multicolumn{1}{c}{\textbf{60}} \\ \midrule
\multicolumn{7}{c}{\textbf{MIND}}                                                                                                                                                                                                                      \\ \midrule
UNBERT                                     & 37.20                           & 38.45                           & 39.49                           & 39.66                           & 40.53                           & 40.30                           \\
MINER                                      & 38.56                           & 39.81                           & 40.14                           & 40.35                           & 40.30                           & 40.65                           \\
UniTRec                                    & 37.55                           & 39.01                           & 39.52                           & 39.69                           & 40.13                           & 40.28                           \\
SPAR                                       & \textbf{38.58}                           & \textbf{39.88}                           & \textbf{40.78}                           & \textbf{40.93}                           & \textbf{41.61}                           & \textbf{41.42}                           \\ \midrule
\multicolumn{7}{c}{\textbf{Goodreads}}                                                                                                                                                                                                                 \\ \midrule
UNBERT                                     & 73.93                           & 74.16                           & 73.68                           & 74.93                           & 75.55                           & 75.21                           \\
MINER                                      & \textbf{73.98}                           & \textbf{74.42}                           & \textbf{74.72}                           & \textbf{75.42}                           & 74.74                           & 74.64                           \\
UniTRec                                    & 72.84                           & 73.28                           & 72.88                           & 73.68                           & 73.67                           & 73.46                           \\
SPAR                                       & 72.04                           & 73.43                           & 74.62                           & 75.39                           & \textbf{75.71}                           & \textbf{75.20}                           \\ \bottomrule
\end{tabular}
\caption{Complete MRR results from investigating the effects of user engagement history length.}\label{tab:history_leng_mrr}
\end{table}

\begin{table}[h]
\centering
\scriptsize
\begin{tabular}{@{}lllllll@{}}
\toprule
\multicolumn{1}{c}{\textbf{History leng.}} & \multicolumn{1}{c}{\textbf{10}} & \multicolumn{1}{c}{\textbf{20}} & \multicolumn{1}{c}{\textbf{30}} & \multicolumn{1}{c}{\textbf{40}} & \multicolumn{1}{c}{\textbf{50}} & \multicolumn{1}{c}{\textbf{60}} \\ \midrule
\multicolumn{7}{c}{\textbf{MIND}}                                                                                                                                                                                                                      \\ \midrule
UNBERT                                     & 32.14                           & 33.45                           & 34.38                           & 34.51                           & 35.30                           & 35.08                           \\
MINER                                      & 33.27                           & 34.47                           & 34.71                           & 34.99                           & 35.03                           & 35.24                           \\
UniTRec                                    & 31.96                           & 33.44                           & 33.90                           & 34.37                           & 34.72                           & 35.07                           \\
SPAR                                       & \textbf{33.58}                           & \textbf{34.81}                           & \textbf{35.46}                           & \textbf{35.72}                           & \textbf{36.22}                           & \textbf{36.16}                           \\ \midrule
\multicolumn{7}{c}{\textbf{Goodreads}}                                                                                                                                                                                                                 \\ \midrule
UNBERT                                     & 52.93                           & 53.68                           & 53.97                           & 54.16                           & 54.96                           & 55.18                           \\
MINER                                      & \textbf{54.12 }                          & \textbf{54.21}                           & 54.45                           & 54.72                           & 54.68                           & 54.31                           \\
UniTRec                                    & 51.70                           & 52.41                           & 52.47                           & 52.97                           & 53.08                           & 53.02                           \\
SPAR                                       & 51.57                           & 54.10                           & \textbf{54.93}                           & \textbf{55.05}                           & \textbf{55.31}                           & \textbf{55.30}                           \\ \bottomrule
\end{tabular}
\caption{Complete ndcg@5 results from investigating the effects of user engagement history length.}\label{tab:history_leng_ndcg5}
\end{table}

\begin{table}[h]
\centering
\scriptsize
\begin{tabular}{@{}lllllll@{}}
\toprule
\multicolumn{1}{c}{\textbf{History leng.}} & \multicolumn{1}{c}{\textbf{10}} & \multicolumn{1}{c}{\textbf{20}} & \multicolumn{1}{c}{\textbf{30}} & \multicolumn{1}{c}{\textbf{40}} & \multicolumn{1}{c}{\textbf{50}} & \multicolumn{1}{c}{\textbf{60}} \\ \midrule
\multicolumn{7}{c}{\textbf{MIND}}                                                                                                                                                                                                                      \\ \midrule
UNBERT                                     & 38.63                           & 39.94                           & 40.61                           & 41.05                           & 41.69                           & 41.86                           \\
MINER                                      & 39.63                           & 40.91                           & 41.39                           & 41.57                           & 41.72                           & 41.86                           \\
UniTRec                                    & 38.23                           & 40.03                           & 40.57                           & 41.00                              & 41.36                           & 41.77                           \\
SPAR                                       & \textbf{40.19}                           & \textbf{41.20}                            & \textbf{42.00}                              & \textbf{42.33}                           & \textbf{42.73}                           & \textbf{42.77}                           \\ \midrule
\multicolumn{7}{c}{\textbf{Goodreads}}                                                                                                                                                                                                                 \\ \midrule
UNBERT                                     & 66.39                           & 66.87                           & 67.2                            & 67.78                           & 68.19                           & 67.78                           \\
MINER                                      & \textbf{67.34}                           & 67.61                           & 67.67                           & 67.93                           & 67.71                           & 67.70                            \\
UniTRec                                    & 65.84                           & 66.11                           & 66.04                           & 66.29                           & 66.24                           & 66.28                           \\
SPAR                                       & 65.56                           & \textbf{67.62}                           & \textbf{68.29}                           & \textbf{68.68}                           & \textbf{68.67}                           & \textbf{68.75}                           \\ \bottomrule
\end{tabular}
\caption{Complete ndcg@10 results from investigating the effects of user engagement history length.}\label{tab:history_leng_ndcg10}
\end{table}

\end{document}